# Density functional analysis of the simultaneous charge/spin order and the associated Cu-Fe intersite charge transfer in LaCu$_3$Fe$_4$O$_{12}$

Changhoon Lee, Erjun Kan and Myung-Hwan Whangbo*

Department of Chemistry, North Carolina State University, Raleigh, North Carolina 27695-8204, USA

**Abstract**

LaCu$_3$Fe$_4$O$_{12}$ undergoes a phase transition at 393 K involving the spin order (SO) of the Fe sites and the charge order (CO) resulting in Cu-Fe intersite charge transfer. On the basis of density functional calculations, we show that this simultaneous CO/SO phenomenon is a consequence of two electronic effects, namely, the competition between the Fe-Fe and Fe-Cu antiferromagnetic spin exchange and the dependence of the width of the Fe d-block bands on the SO of the Fe atoms.

Magnetic oxides of transition-metal elements may undergo spin order (SO), orbital order and/or charge order (CO). As found for $LuFe_2O_4$ [1-3] and $Fe_2OBO_3$ [4,5], a system undergoing CO may also exhibit SO. The CO and SO take place at widely different temperatures in $LuFe_2O_4$ and $Fe_2OBO_3$ (below 330 [1-3] and 240 K [3], respectively, for $LuFe_2O_4$, and 340 [4] and 155 K [5], respectively, for $Fe_2OBO_3$), showing that their CO and SO are not coupled. Recently, $LaCu_3Fe_4O_{12}$ was found to undergo a phase transition at $T_p$ = 393 K [6] such that the Cu and Fe atoms exist as $Cu^{3+}$ and $Fe^{3+}$ ions below $T_p$, but as $Cu^{2+}$ and $Fe^{3.75+}$ ions above $T_p$. This CO involves the transfer of three electrons per formula unit (FU) between the Cu and Fe sites. The magnetic susceptibility of $LaCu_3Fe_4O_{12}$ suggests that it is paramagnetic (PM) above $T_p$ but undergoes an antiferromagnetic (AFM) ordering between the $Fe^{3+}$ ions below $T_p$ [6]. The occurrence of both CO and SO at the same temperature suggests that they are strongly coupled. In this Letter we explore the origin of this simultaneous CO/SO and the associated Cu-Fe intersite charge transfer in $LuFe_2O_4$.

In understanding the reason for the simultaneous CO/SO in $LaCu_3Fe_4O_{12}$, two observations are important. One is the implication of the AFM ordering below $T_p$ concerning the spin exchange between nearest-neighbor (NN) Fe and Cu atoms. In the crystal structure of $LaCu_3Fe_4O_{12}$, the $FeO_6$ octahedra and $CuO_4$ square planes share their corners (Fig. 1a,b) such that the Fe atoms form a simple cubic lattice with the Cu and La atoms occupying the centers of the $Fe_8$ cubes (Fig. 1c-e) [6]. Suppose that the Cu atoms exist as magnetic $Cu^{2+}$ ions, the spin exchange between NN Fe and Cu atoms is AFM, and so is that between NN Fe atoms. If the cubic Fe lattice has an AFM ordering (Fig. 2a), the exchange interactions between NN Fe and Cu atoms are frustrated (Fig. 2b,c) [7]. This spin frustration can be removed by converting the $Cu^{2+}$ ions nonmagnetic $Cu^{3+}$ ions. When the cubic Fe lattice has a ferromagnetic (FM) arrangement (Fig. 2d), the AFM coupling between NN Fe and Cu atoms leads to energy stabilization (Fig. 2e). The second observation to notice, for a transition-metal magnetic solid, is that the width of its d-block bands depends on the nature of its ordered spin arrangement; the width is wider in the FM than in the AFM state [8]. Since all Fe atoms of $LaCu_3Fe_4O_{12}$ are interconnected via Fe-O-Fe linkages, its Fe d-block bands should be wider for the FM than for the AFM arrangement of the Fe spins. In contrast, the Cu atoms of $LaCu_3Fe_4O_{12}$ are isolated, so the Cu $x^2$-$y^2$ band (i.e., the highest-occupied Cu d-block band) should be narrow. Consequently, electron transfer would occur from the Fe to the Cu sites if the Fe d-block bands are wide with their top lying above the Cu $x^2$-$y^2$ band, while the reverse electron transfer would occur if the Fe d-block bands are narrow with their top lying below the Cu $x^2$-$y^2$ band. In the following we verify this suggestion on the basis of density functional theory (DFT) calculations.

Our DFT calculations for $LaCu_3Fe_4O_{12}$ employed the projector augmented wave method [9] coded in the Vienna *ab initio* simulation package [10] with the generalized gradient approximation (GGA) for the exchange and correlation correction [11], the plane wave cut off energy of 400 eV, and the sampling of the irreducible Brillouin zone with 32 k-points. For the crystal structures of $LaCu_3Fe_4O_{12}$ at 100, 300 and 450 K [6], our GGA calculations predict nonzero spin moments on both Fe and Cu (e.g., $\mu_{Fe}$ = 3.00 $\mu_B$ and $\mu_{Cu}$ = 0.17 $\mu_B$ for the 100 K structure), and metallic properties for $LaCu_3Fe_4O_{12}$. $LaCu_3Fe_4O_{12}$ is metallic above $T_p$ but insulating below $T_p$ [6]. To predict the insulating property of $LaCu_3Fe_4O_{12}$ below $T_p$, one needs to take into consideration the electron correlation in the Fe 3d and Cu 3d states as well as ordered spin states of $LaCu_3Fe_4O_{12}$. Thus, we employed the GGA plus onsite repulsion (GGA+U) calculations [12] with $U_{Fe}$ = 3 eV and $U_{Cu}$ = 5 eV to take into consideration the electron correlation in the Fe 3d and Cu 3d states, and considered three ordered spin states of $LaCu_3Fe_4O_{12}$; (1) the FM/FM state in which the NN Fe atoms have an FM coupling, and so do the NN Fe and Cu atoms (Fig. 2f), (2) the FM/AFM state in which the NN Fe atoms have an FM coupling, but the NN Fe and Cu atoms have an AFM coupling (Fig. 2e), and (3) the AFM/NM state in which the NN Fe atoms have an AFM coupling, and the Cu atoms are nonmagnetic (NM) (Fig. 2a).

Table 1a summarizes the relative energies (in meV per two FUs) of the FM/FM, FM/AFM and AFM/NM states calculated for the 100, 300 and 450 K structures. The calculated magnetic moments of the Fe and Cu atoms are summarized in Table 1b, and the PDOS plots calculated for the Fe 3d and Cu 3d states in Fig. 3. In Table 1b and Fig. 3, the results for the FM/FM state are not included because they are quite similar to those for the FM/AFM state. Further, the PDOS plots for the 300 and 450 K structures are similar to those for the 100 K structure, and hence are not shown in Fig. 3. Several important observations emerge from these results:

(a) For the crystal structures above and below $T_p$, the Cu atoms exist as magnetic $Cu^{2+}$ ions for the FM coupling of the Fe spins, but as nonmagnetic $Cu^{3+}$ ions for the AFM coupling of the Fe spins (Table 1).

(b) $LaCu_3Fe_4O_{12}$ is metallic for the FM coupling of the Fe spins, but are insulating for the AFM coupling of the Fe spins (Fig. 3). The Fe d-block bands are wider for the FM than for the AFM coupling of the Fe spins (Fig. 3).

(c) For the AFM coupling of the Fe spins, the top of the majority-spin Fe d-block bands lies below the Fermi level and is separated from the bottom of the minority-spin Fe d-block bands by a band gap (Fig. 3a). The up- and down-spin Cu $x^2$-$y^2$ bands lie above the Fermi level and are separated from the occupied Cu d-block bands by a band gap (Fig. 3a). This gives rise to the oxidation assignments $Fe^{3+}$ and $Cu^{3+}$.

(d) For the FM coupling of the Fe spins, the top of the majority-spin Fe d-block bands lies above the Fermi level (Fig. 3b), while the Cu $x^2$-$y^2$ bands are spin polarized such that the minority-spin $x^2$-$y^2$ band lies above the Fermi level and is separated from the occupied Cu d-block bands by a band gap (Fig. 3b). This leads to the oxidation assignments $Fe^{3.75+}$ and $Cu^{2+}$.

(e) For the FM coupling of the Fe spins, the NN Fe and Cu atoms prefer to have an AFM coupling (Table 1a). For the structures below $T_p$, the AFM/NM state is more stable than the FM/AFM and FM/FM states (Table 1a). For the structure above $T_p$, the AFM/NM state is less stable that the FM/AFM state (Table 1a).

The above findings support our suggestion that the CO and SO are intimately coupled in $LaCu_3Fe_4O_{12}$. In discussing the magnetic structure of $LaCu_3Fe_4O_{12}$ above $T_p$, however, we considered the 3D FM arrangement of the Fe spins (Fig. 2e,f). Since $LaCu_3Fe_4O_{12}$ is PM above $T_p$ [6], it is necessary to examine how the electronic structure of $LaCu_3Fe_4O_{12}$ might be affected by the randomness in the Fe spin arrangements above $T_p$. Thus, we considered two additional ordered spin arrangements for the 450 K structure, namely, the two-dimensional (2D) FM arrangement (Fig. 2g) in which layers of ferromagnetically ordered Fe spins are antiferromagnetically coupled and the one-dimensional (1D) FM arrangement (Fig. 2h) in which chains of ferromagnetically ordered Fe spins are antiferromagnetically coupled. The PDOS plots of the Fe and Cu d-block bands calculated for the 2D and 1D FM arrangements of the 450 K structure, presented in Fig. 4, reveal that the top of their Fe d-block bands lies above the Fermi level with the minority-spin Cu $x^2$-$y^2$ band located above the Fermi level. As in the case of the 3D FM arrangement, therefore, the Fe and Cu atoms of the 2D and 1D FM arrangements are present as $Fe^{3.75+}$ and $Cu^{2+}$ ions, respectively ($\mu_{Fe}$ = 3.82 $\mu_B$ and $\mu_{Cu}$ = 0.54 $\mu_B$ for the 2D FM arrangement, and $\mu_{Fe}$ = 3.81 $\mu_B$ and $\mu_{Cu}$ = 0.49 $\mu_B$ for the 1D FM arrangement). The 2D and 1D FM arrangements are calculated to be substantially less stable than the 3D FM arrangement (i.e., the FM/AFM state) by 683.0 and 734.8 meV per two FUs, respectively. This is expected because the Fe-Cu spin exchange interactions are frustrated in the 2D and 1D FM arrangements (Fig. 2g,h) but are antiferromagnetically coupled in the 3D FM arrangement (Fig. 2e). Thus, to the PM state of $LaCu_3Fe_4O_{12}$ above $T_p$, the 3D FM arrangement should contribute more strongly than do the 2D and 1D FM arrangements. When the 3D FM spin arrangement of the Fe spins is partially broken by thermal agitation above $T_p$, the Fe and Cu atoms are expected to be present as $Fe^{3.75+}$ and $Cu^{2+}$ ions, respectively.

Finally, we examine the SO of $LaCu_3Fe_4O_{12}$ in terms of the spin exchange $J_{Fe-Fe}$ between NN Fe atoms and the spin exchange $J_{Fe-Cu}$ between NN Fe and Cu atoms. The $J_{Fe-Cu}$ interactions occur more than do $J_{Fe-Fe}$ interactions by a factor of two. To evaluate $J_{Fe-Fe}$ and $J_{Fe-Cu}$ we consider the spin exchange interaction energies of the ordered spin states of $LaCu_3Fe_4O_{12}$ using the spin Hamiltonian

$$\hat{H} = \sum_{i<j} J_{ij}\hat{S}_i \cdot \hat{S}_j$$

where $J_{ij}$ is the spin exchange between the sites i and j (i.e., $J_{ij}$ = $J_{Fe-Fe}$ or $J_{Fe-Cu}$). Two states, AFM/NM and FM/NM, were for the calculation of $J_{Fe-Fe}$. The AFM/NM state was already discussed. In the FM/NM state, the NN Fe atoms have an FM coupling while the Cu atoms are NM. The Cu atoms become magnetic for the FM coupling of the Fe spins, as long as the nonzero $U_{Cu}$ allows spin polarization on Cu. To prevent this, we carried out GGA+U calculations for the AFM/NM and FM/NM states using $U_{Fe}$ = 3 eV and $U_{Cu}$ = 0. For the 100, 300 and 450 K structures, the AFM/NM state is calculated to be more stable than the FM/NM state by 1355.3, 1231.9 and 714.8 meV per two FUs, respectively. The total spin exchange energies per two FUs of the two spin states are written as E(FM/NM) = $+24S_{Fe}^2 J_{Fe\text{-}Fe}$, and E(AFM/NM) = $-24S_{Fe}^2 J_{Fe\text{-}Fe}$, where $S_{Fe}$ is the spin on $Fe^{3+}$ (i.e., $S_{Fe}$ = 5/2) by using the energy expressions for spin dimmers [13]. Thus, by mapping these energy difference to the corresponding energy difference of the GGA+U calculations, we obtain the $S_{Fe}^2 J_{Fe-Fe}$ value (Table 2). To estimate $J_{Cu\text{-}Fe}$, we employed the FM/FM and FM/AFM states (Table 1a). When the FM coupling of the Fe spins, the Fe and Cu atoms exist as $Fe^{3.75+}$ and $Cu^{2+}$ ions, respectively. Formally, the high-spin $Fe^{3.75+}$ ion has 4.25 unpaired spins. We treat each $Fe^{3.75+}$ ion as having four unpaired spins (i.e., $S_{Fe}$ = 2), and let the $J_{Fe\text{-}Cu}$ value absorb the effect of neglecting 0.25 unpaired spin. Then, the total spin exchange energies of the FM/FM and AF/AFM states, per two FUs, is written as E(FM/FM) = $+48 S_{Fe} S_{Cu} J_{Cu\text{-}Fe} + 24 S_{Fe}^2 J_{Fe\text{-}Fe}$, and E(FM/AFM) = $-48 S_{Fe} S_{Cu} J_{Cu\text{-}Fe} + 24 S_{Fe}^2 J_{Fe\text{-}Fe}$, where $S_{Cu}$ = 1/2 for the $Cu^{2+}$ ions and $S_{Fe}$ = 2 for the $Fe^{3.75+}$ ions. Thus, from the energy difference between FM/AFM and FM/FM states in Table 1a, we obtain $S_{Fe}S_{Cu}J_{Fe\text{-}Cu}$ (Table 2). Table 2 reveals that the Fe-Fe and the Fe-Cu exchange interactions are AFM, the spin exchange $S_{Fe}^2 J_{Fe-Fe}$ is stronger below $T_p$ than above $T_p$ by a factor of ~2, but the opposite is true for the spin exchange $S_{Fe}S_{Cu}J_{Fe\text{-}Cu}$. At 450 K, $S_{Fe}^2 J_{Fe-Fe}$ and $S_{Fe}S_{Cu}J_{Fe\text{-}Cu}$ are nearly comparable in magnitude, so that the FM/AFM state is more stable than the AFM/NM state because there occur twice more $S_{Fe}S_{Cu}J_{Fe\text{-}Cu}$ interactions than do the $S_{Fe}^2 J_{Fe-Fe}$ interactions. Below $T_p$, $S_{Fe}^2 J_{Fe-Fe}$ is stronger than $S_{Fe}S_{Cu}J_{Fe\text{-}Cu}$ by a factor of ~4, which makes it energetically favorable for $LaCu_3Fe_4O_{12}$ to adopt the AFM/NM state in which the $S_{Fe}^2 J_{Fe-Fe}$ interactions are maximized by converting the $Cu^{2+}$ ions to $Cu^{3+}$ ions and hence removing the spin frustration associated with the $S_{Fe}S_{Cu}J_{Fe\text{-}Cu}$ interactions.

In summary, the simultaneous occurrence of CO and SO and the associated Cu-Fe intersite charge transfer in LaCu$_3$Fe$_4$O$_{12}$ originate from two electronic factors, i.e., the competition between the Fe-Fe and Fe-Cu AFM spin exchange interactions and the dependence of the width of the Fe d-block band on the magnetic order between the Fe atoms.

**Acknowledgements**

The research was supported by the Office of Basic Energy Sciences, Division of Materials Sciences, U.S. Department of Energy, under Grant No. DE-FG02- 86ER45259, and by the National Energy Research Scientific Computing Center under Contract No. DE-AC02-05CH11231.

Table 1. Results of the GGA+U calculations for the 100, 300 and 450 K structures of LaCu$_3$Fe$_4$O$_{12}$ with U$_{Fe}$ = 3.0 eV and U$_{Cu}$ = 5.0 eV

(a) Relative energies (in meV per two FUs) of the FM/FM, FM/AFM and AFM/NM spin states

|        | 100 K  | 300 K  | 450 K  |
|--------|--------|--------|--------|
| FM/FM  | 1562.5 | 1526.0 | 1347.6 |
| FM/AFM | 949.1  | 884.9  | 176.0  |
| AFM/NM | 0.0    | 22.3   | 763.3  |

(b) Magnetic moments of the Cu and Fe atoms (in µ$_B$) in the FM/AFM and AFM/NM states

|         | 100 K  |        | 300 K  |        | 450 K  |        |
|---------|--------|--------|--------|--------|--------|--------|
|         | FM/AFM | AFM/NM | AM/AFM | AFM/NM | FM/AFM | AFM/NM |
| µ$_{Fe}$ | 3.81   | 4.00   | 3.78   | 4.00   | 3.74   | 3.96   |
| µ$_{Cu}$ | 0.47   | 0.00   | 0.52   | 0.00   | 0.57   | 0.00   |

Table 2. Spin exchange parameters $S_{Fe}^2 J_{Fe\text{-}Fe}$ and $S_{Fe}S_{Cu}J_{Fe\text{-}Cu}$ (in meV) calculated for the 100, 300 and 450 K structures of $LaCu_3Fe_4O_{12}$ from GGA+U calculations

|  | 100 K | 300 K | 450 K |
|---|---|---|---|
| $S_{Fe}^2 J_{Fe\text{-}Fe}$ | 28.25 | 25.66 | 14.88 |
| $S_{Fe}S_{Cu}J_{Fe\text{-}Cu}$ | 6.39 | 6.68 | 12.20 |

**Figure captions**

Figure 1.   Arrangement of the $FeO_6$ octahedra and $CuO_4$ square planes and that of the La, Cu and Fe atoms in the crystal structure of $LaCu_3Fe_4O_{12}$: (a) An $FeO_6$ octahedron sharing corners with six $CuO_4$ square planes. (b) A $CuO_4$ square plane sharing corners with eight $FeO_6$ octahedra. (c) A $Fe_8Cu$ cube. (d) A perspective view of the 3D arrangement of the La, Cu and Fe atoms. (e) A $Fe_8La$ cube capped with six $CuO_4$ square planes. The La, Cu, Fe and O atoms are represented by cyan, blue, red and white circles, respectively.

Figure 2.   Ordered spin arrangements of the Fe and Cu atoms in $LaCu_3Fe_4O_{12}$: (a) The AFM arrangement between the NN Fe atoms in a $Fe_8$ cube. (b, c) The spin frustration between the NN Fe and Cu atoms in an antiferromagnetically coupled $Fe_8$ cube. (d) The FM arrangement between the NN Fe atoms in a $Fe_8$ cube. (e) The AFM arrangement between the NN Fe and Cu atoms in a ferromagnetically coupled $Fe_8$ cube. (f) The FM arrangement between the NN Fe and Cu atoms in a ferromagnetically coupled $Fe_8$ cube. (g) The 2D FM arrangement in which layers of ferromagnetically ordered Fe spins are antiferromagnetically coupled. (h) The 1D FM arrangement in which chains of ferromagnetically ordered Fe spins are antiferromagnetically coupled.

Figure 3.   PDOS plots of the Fe 3d and Cu 3d states obtained for (a) the AFM/NM and (b) the FM/AFM states of the 100 K structure of $LaCu_3Fe_4O_{12}$ from GGA+U calculations with $U_{Fe}$ = 3 eV and $U_{Cu}$ = 5 eV. The PDOS plots for the Cu $x^2$-$y^2$ bands are shaded in green.

Figure 4. PDOS plots of the Fe 3d and Cu 3d states obtained for (a) the 2D FM and (b) the 1D FM arrangements of the 450 K structure of LaCu$_3$Fe$_4$O$_{12}$ from GGA+U calculations with U$_{Fe}$ = 3 eV and U$_{Cu}$ = 5 eV. The PDOS plots for the Cu $x^2$-$y^2$ bands are shaded in green.

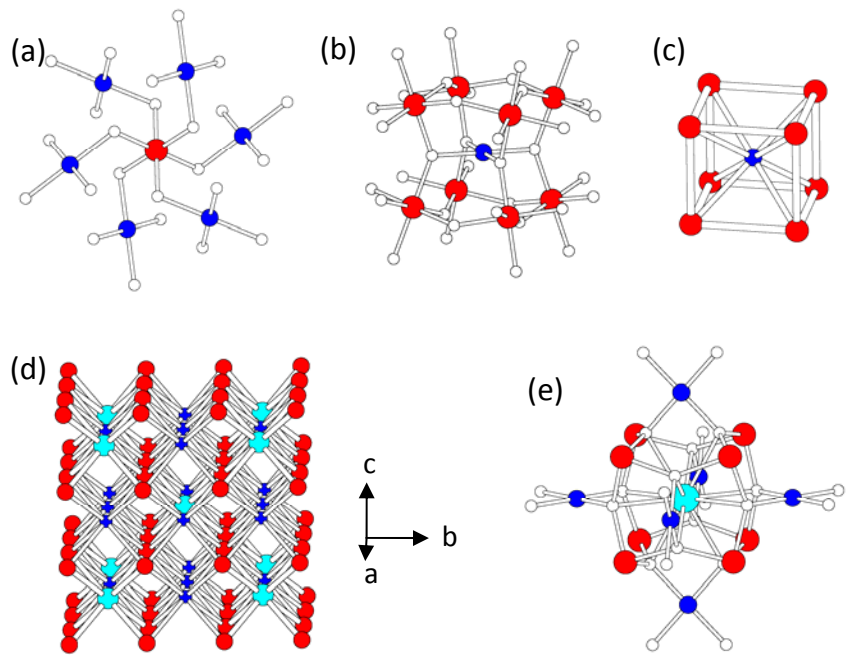

Figure 1.

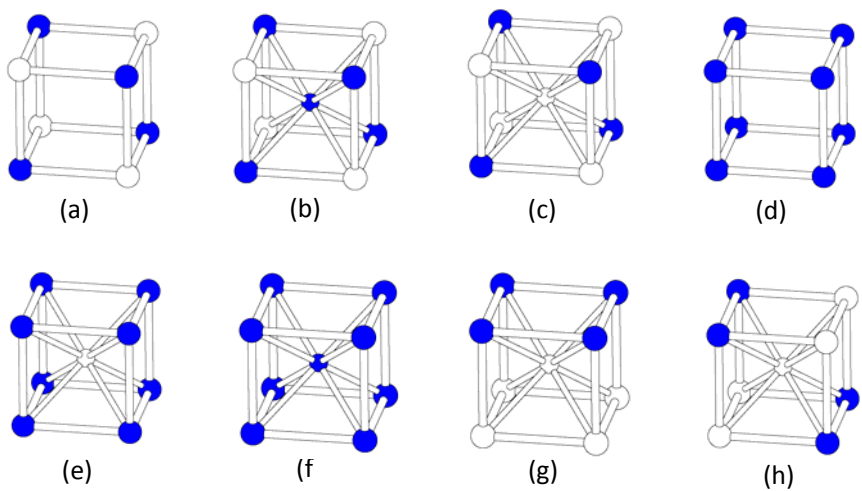

Figure 2.

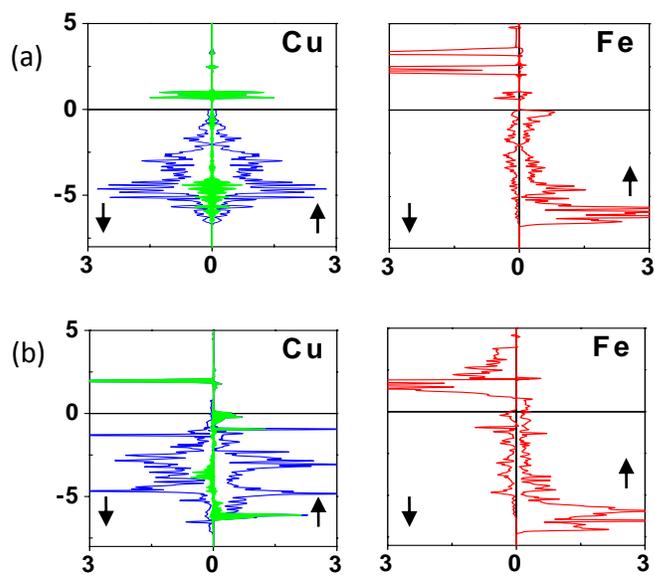

Figure 3.

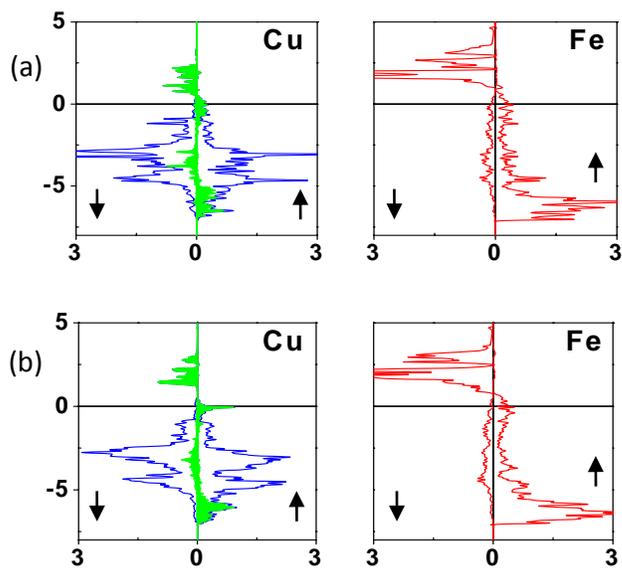

Figure 4.